\documentclass{article}

\usepackage{moreverb,url}

\usepackage{graphicx}   %for subplots
\usepackage{subfig} %for plots
\usepackage{natbib}
\usepackage{float} % to force table stay in position
\usepackage{chngcntr}
\usepackage{amsmath}
\usepackage{amsfonts}
\usepackage{adjustbox} %to trim pics
\usepackage[colorlinks,bookmarksopen,bookmarksnumbered,citecolor=red,urlcolor=red]{hyperref}

\newcommand\BibTeX{{\rmfamily B\kern-.05em \textsc{i\kern-.025em b}\kern-.08em
T\kern-.1667em\lower.7ex\hbox{E}\kern-.125emX}}

\begin{document}

\title{A Spatially Discrete Approximation to Log-Gaussian Cox Processes for Modelling Aggregated Disease Count Data}

\author{Olatunji Olugoke Johnson, Peter Diggle and \\ Emanuele Giorgi \\ \\
CHICAS, Lancaster Medical School, \\ Lancaster University, Lancaster, UK}

\maketitle

\begin{abstract}
In this paper, we develop a computationally efficient discrete approximation to log-Gaussian Cox process (LGCP) models for the analysis of spatially aggregated disease count data. Our approach overcomes an inherent limitation of  spatial models based on Markov structures,
namely that each such model is tied to a specific partition of the study area, and allows for spatially continuous prediction. We compare the predictive performance of our modelling approach with LGCP through a simulation study and an application to primary biliary cirrhosis incidence data in Newcastle-Upon-Tyne, UK. Our results suggest that when disease risk is assumed to be a spatially continuous process, the proposed approximation to LGCP provides reliable estimates of disease risk both on spatially continuous and aggregated scales. The proposed
methodology is implemented in the open-source R package \texttt{SDALGCP}. \\

\end{abstract}

\section{Introduction}
%brief background

In this paper our concern is to make inference on a spatially continuous disease risk surface using aggregated counts of reported disease cases, say $y_i$, over regions $\mathcal{R}_i$ forming a partition of a geographical area of interest $A$. In this context, information on risk factors and on the population at risk may also be available, 
possibly at different spatial scales. We shall denote these by $d(x)$ and
$m(x)$, respectively, when available on a spatially continuous scale, and by $d_i$ and $m_i$ when they are spatially aggregated. \par
Existing methods from small area estimation (SAE) only allow spatial prediction at the aggregated level of the regions $\mathcal{R}_i$ and are usually based on a Gaussian Markov random field (GMRF) structure. \citep{besag1974spatial, rue2005} Typically,
non-zero elements of the precision matrix of a GMRF are restricted to
contiguous pairs of the $\mathcal{R}_i$. Hence, the formulation and interpretation of a GMRF is tied to the specific partition of $A$,
which will usually have been drawn up for administrative, historical, 
or other reasons unrelated to the disease aetiology. 
The use of such models also becomes impractical when the spatial units $\mathcal{R}_i$ change over time. Wall \cite{wall2004close} points out
that the use of GMRFs is especially problematic when dealing with irregular geometries, which can induce counter-intuitive forms for the correlation structure between variables associated with
the $\mathcal{R}_i$. \par
The geostatistical paradigm, unlike SAE, treats disease risk as a spatially continuous phenomenon irrespective of the data-format. Diggle et al\cite{diggle2013spatial} argue that the analysis of spatially aggregated counts can be regarded as a special case of the class of geostatistical problems and propose to model the $y_i$ as an aggregated realisation of a Log-Gaussian Cox process (LGCP). Unlike GMRFs, LGCPs allow for prediction of disease risk at any spatial scale, while avoiding the ecological fallacy.\citep{wakefield2006health} However, fitting of LGCP models using the aggregated counts $y_i$ is computationally demanding due to the iterative imputation of the unobserved locations for each reported case within a region $\mathcal{R}_i$.\citep{li2012log} \par
In this paper, our objective is to develop a computationally efficient approximation to LGCPs in order to predict disease risk at any desired spatial scale. We argue that this provides a more realistic alternative to GMRF models when LGCPs are not computationally feasible, and can also be used as an exploratory tool in order to inform more complex modelling approaches based on LGCPs. \par

In Section \ref{sec2} of the paper, we review existing methods for modelling spatially aggregated disease counts.  In Section \ref{sec:sdlgcp}, we develop a computationally efficient spatially discrete approximation to LGCP models. In Section \ref{sec:simulation} we carry out a simulation study to investigate the predictive performance of the proposed approximation and compare this with an exact fitting algorithm for LGCP models. In Section \ref{sec:application} we show an application of the method to a data-set on primary biliary cirrhosis (PBC) incidence in Newcastle, UK. Section \ref{sec6} is a concluding discussion on the advantages and limitations of the proposed approach.

The method has been implemented in the open-source R package \texttt{SDALGCP},\citep{SDALGCP} available from the Comprehensive R Network Archive. The R code for reproducing the results of Section \ref{sec:application} is available as supplementary material.

\section{Existing methods for modelling spatially aggregated disease counts data} \label{sec2}

\subsection{Gaussian Markov random field models}
\label{subsec:discrete}
Let $Y_{i}$ denote the reported disease count in region
$\mathcal{R}_{i}$. Conditionally on a zero-mean Gaussian 
process $S= (S_1,\ldots,S_n)$, assume that the $Y_{i}$ are mutually independent Poisson random variables with expectations
\begin{equation}
\label{eq:carmodel}
\lambda_{i} = m_{i} \exp\{d_{i}^\top \beta + S_{i}\}, i=1,\ldots,n
\end{equation}
where $\beta$ is a vector of regression coefficients and $m_i$ is the population count or a standardised expectation of the number of cases, taking into account the demographics of the population in subregion $\mathcal{R}_i$ but assuming that risk is otherwise spatially homogeneous. Spatially discrete models are then developed by specifying the precision matrix for the Gaussian process $S$. Here, we focus on the two most commonly used formulations for $S$, namely the conditional autoregressive (CAR) \citep{leroux2000estimation} and intrinsic conditional autoregressive (ICAR) \citep{besag1991bayesian} models. \par
Let $i \sim j$ be a shorthand notation for ``$\mathcal{R}_i$ and $\mathcal{R}_j$ are neighbours''. A CAR model then assumes that
\begin{equation}
\label{eq:car}
S_{i} | S_{-i} \sim N\left(\rho_{c} \sum_{j \sim i} c_{ij} S_{j}, \tau_i^2\right),
\end{equation}
where $S_{-i} = \{S_{j}: j\neq i \}$, $\rho_{c}$ is the spatial dependence parameter and $c_{ij}$ are known quantities such that $c_{ij} \neq 0$ if and only if $j\sim i$ and $j \neq i$. It follows from
Brook's Lemma \citep{brook1964} and the Hammersley-Clifford Theorem \citep{besag1974spatial} that the joint distribution of $S$ is a multivariate zero-mean Gaussian distribution with covariance matrix
\begin{equation}
\label{eq:car2}
(I- \rho_c C)^{-1} \tilde{D},
\end{equation}
where $\tilde{D}=\{\tau_{1}^2,\ldots,\tau_{n}^2\}$, while the specification of $C$ is generally tied to the specific arrangement of the partition of the region of interest. The most common approach is to set $c_{ij} =1$ if $j \sim i$ and 0 otherwise. The matrix in \eqref{eq:car2} is then a valid covariance matrix if $\xi_{max}^{-1} < \rho_{c} < \xi_{min}^{-1}$ \citep[pg. 472]{cressie1993statistics}, where $\xi_{min}$ and $\xi_{max}$ are the minimum and maximum eigenvalues of $C$, respectively. Scaling of the matrix $C$ so as to obtain a weighted average of the $S_{j}$ in \eqref{eq:car} also implies that $-1<\rho_{c}<1$. \par
The ICAR model is a special case of the CAR model when $\rho_c =1$ in \eqref{eq:car}. Although this leads to an improper distribution for $S$ because of the singularity of its covariance matrix, the associated 
conditional distribution of $S$ given $Y$ is proper.

\subsection{Log-Gaussian Cox process models}
\label{subsec:lgcp}

A spatial point process is a stochastic mechanism that generates a 
countable set of events $x_i \in \mathbb{R}^2$. The class of inhomogeneous Poisson processes with intensity $\lambda(x)$
is defined by the following postulates.
\begin{enumerate}
	\item The number of events, $N(\mathcal{A})$, in any planar
	region $\mathcal{A} \subset \mathbb{R}^2$ follows a Poisson
	distribution with mean $\int_{\mathcal{A}} \lambda(x) dx$.
	\item Conditionally on $N(\mathcal{A})$, each event 
	in $\mathcal{A}$
	is an independent random sample from a distribution on $\mathcal{A}$ with probability density function proportional to $\lambda(x)$.
\end{enumerate}  \par

A Cox process \citep{cox1955some} is defined by a non-negative valued
stochastic process $\Lambda(x)$ such that,
conditional on a realisation of $\Lambda(x)$, the process
is an inhomogenous Poisson process with intensity $\Lambda(x)$. If we assume that $\log\{\Lambda(x)\} = S(x)$ is a Gaussian process, we obtain the log-Gaussian Cox process (LGCP); for more details on the theoretical properties of LGCPs, see Moller et al\cite{moller1998log} \par

Diggle \cite{diggle2013spatial} develop a modelling framework for aggregated disease count data using LGCPs. They assume that, 
conditionally on $S(x)$, the $Y_{i}$ are mutually independent Poisson variables with means 
\begin{equation}
\label{eq:lgcpmodel}
\int_{\mathcal{R}_{i}} m(x) \exp\{d(x)^\top \beta + S(x) \} \: dx,
\end{equation}
where $d(x)$ is a vector of covariates at location $x$ with associated regression coefficients $\beta$.

A first notable difference between \eqref{eq:carmodel} and \eqref{eq:lgcpmodel} is that the latter uses spatially continuous information on the distribution of the expected cases, $m(x)$, hence, unlike \eqref{eq:carmodel}, avoids the questionable assumption of a homogeneous distribution of the population at risk within $\mathcal{R}_i$. However, population density is often only available in the form of small-area population counts, implying a
piece-wise constant surface $m(x)$. Note, however, that modelled spatially continuous maps for population density have been made freely available; see, for example, \texttt{sedac.ciesin.columbia.edu/data/collection/gpw-v4}.

Furthermore, unlike the spatially discrete models described in the previous section, LGCP is not tied to any particular partition of the area of interest and therefore provides a route to a solution to the problem of combining information at multiple spatial scales. However, this is offset by a substantial increase in the computational burden arising from the need to impute the unobserved locations for each of the reported cases within each of the $\mathcal{R}_i$, $i=1,\ldots,n$.\citep{li2012log} In the next section, we circumvent this issue by proposing a spatially discrete approximation to $S(x)$ which allows to model the counts $y_i$ as the realisation of a Poisson log-linear mixed model.\par

\section{A spatially discrete approximation to Log-Gaussian Cox processes}
\label{sec:sdlgcp}
Let $w_i(x)$ be a positive function with domain $\mathcal{R}_i$, such that $\int_{\mathcal{R}_i} w_{i}(x) \: dx = 1$. Using the same notation as in Section \ref{subsec:lgcp}, we approximate the conditional log-intensity of an LGCP as piecewise constant by taking its weighted average over $\mathcal{R}_i$ to give
\begin{eqnarray}
\label{eq:log_intens}
\log\{\Lambda(x)\} &\approx & \int_{\mathcal{R}_{i}} w_i(x) \left[ d(x)^\top \beta^* + S^*(x) \right]\: dx \nonumber \\
&=& \int_{\mathcal{R}_{i}} w_i(x) \: d(x)^\top \beta^* \: dx +  \int_{\mathcal{R}_{i}} w_i(x) \: S^*(x) \: dx \nonumber \\
&=& d_{i}^\top \beta^* + S_{i}^*, \: x \in \mathcal{R}_i,
\end{eqnarray}
where $\beta^*$ is a vector of regression coefficients for the aggregate explanatory variables $d_i$ and $S^*(x)$ is a Gaussian process. 
The rationale for using the weighting function $w_i(x)$ is to account for the potential non-homogeneous distribution of disease cases within a region $\mathcal{R}_i$. For example, a larger number of cases may concentrate in more densely populated areas, thus a natural choice for $w_i(x)$ would be to set this equal to $m(x)/m_i$ with $m_i=\int_{\mathcal{R}_i} m(x) dx$, if $m(x)$ is available. If $m(x)$ is instead unavailable, a pragmatic approach would be to set $w_{i}(x)=1/|\mathcal{R}_i|$.    \par

Following from \eqref{eq:log_intens}, we obtain the following approximation for the conditional mean of the counts $Y_i$
\begin{eqnarray}
\label{eq:approx}
\lambda_{i} =  \int_{\mathcal{R}_{i}} m(x) \Lambda(x) \: dx &\approx &  \int_{\mathcal{R}_{i}} m(x) \exp\left\{d_{i}^{\top} \beta^* + S_{i}^* \right \} \:dx\nonumber \\
&=&  m_{i} \exp\{d_{i}^{\top}\beta^* + S_{i}^*\} \nonumber \\
&=&  m_i \exp\{\eta_i\} \nonumber \\
&=& \mu_{i}.
\end{eqnarray}

The joint distribution of $S^* = (S_{1}^*,\ldots, S^*_{n})$ is multivariate Gaussian with zero mean and covariance function
\begin{equation}
\label{eq:covariance}
\text{Cov}\{S_{i}^*, S_{j}^*\} = \sigma^2 \int_{\mathcal{R}_{i}} \int_{\mathcal{R}_{j}} w_i(x) w_j(x') \: \rho(\|x-x'\|; \phi) \:  dx \: dx',
\end{equation}
where $\| \cdot \|$ is the Euclidean distance and $\rho(\cdot; \phi)$ is the isotropic and stationary covariance function of $S^*(x)$ indexed by the parameter $\phi$. Hence, the resulting model \eqref{eq:approx} falls under the class of generalized linear mixed models. Also, note that the variance of $S_{i}^*$ depends on the size and shape of $\mathcal{R}_i$, with larger regions leading to smaller variances. \par

We now provide further details on the computation of the covariance function in \eqref{eq:covariance}. Among the class of isotropic and stationary covariance functions for $S^*(x)$ in \eqref{eq:approx}, one of the most commonly used is the Mat\'ern covariance function,\citep{stein2012interpolation} which has expression
\begin{equation}
\label{eq:matern}
\text{Cov}\{S^*(x),S^*(x')\} = \frac{\sigma^2}{2^{\kappa-1} \Gamma(\kappa)}\left(\frac{u}{\phi} \right)^\kappa \mathcal{K}_\kappa \left(\frac{u}{\phi} \right),
\end{equation}
where $u=\|x-x'\|$ is the Euclidean distance between any two locations $x$ and $x'$, $\sigma^2$ is the variance, $\phi$ is a scale parameter that regulates the rate at which the spatial correlation decays for increasing distance $u$, $\kappa$ is the shape parameter that determines the differentiability of the process $S$  and $\mathcal{K}_\kappa(\cdot)$  is the modified Bessel function of the second kind of order $\kappa>0$. Estimating $\kappa$ reliably requires a large amount of densely sampled data, which in this context is not available.  As shown by Zhang,\cite{zhang2004inconsistent} not all of the three parameters $\sigma^2$, $\phi$ and $\kappa$ can be consistently estimated under in-fill asymptotics, and in practice this translates to $\kappa$ often being poorly identified. This issue is likely to be further exacerbated in this context. As a pragmatic approach, we then set $\kappa=0.5$ which reduces \eqref{eq:matern} to 
\begin{equation*}
\text{Cov}\{S^*(x),S^*(x')\} = \sigma^2 \exp\{-u/\phi\}
\end{equation*} 
corresponding to a mean-square continuous process. \par

We approximate \eqref{eq:covariance} as a discrete sum over $L_{i}$ and $L_{j}$ randomly chosen points in  $\mathcal{R}_{i}$ and $\mathcal{R}_{j}$ to give
\begin{eqnarray}
\small
\label{eq:IntegralApprox}
\int_{\mathcal{R}_{i}} \int_{\mathcal{R}_{j}} w_i(x) w_j(x') \: \rho(\|x-x'\|; \phi) \: dx \: dx' \approx \nonumber  &&\\  \frac{\sum_{k=1}^{L_{i}} \sum_{k'=1}^{L_{j}} w_i(x_k) w_j(x_{k'}) \: \rho(\|x_k-x_{k'}\|; \phi)}{\sum_{k=1}^{L_{i}} \sum_{k'=1}^{L_{j}} w_i(x_k) w_j(x_{k'})},
\end{eqnarray}
To attain a good spatial coverage of $\mathcal{R}_i$ and $\mathcal{R}_j$, we propose to draw each of the $x_{k}$ and $x_{k'}$ in the above equation using a class of inhibition processes \citep[pp.~110-116]{diggle2013statistical} which combine simple sequential inhibition with rejection sampling. More specifically, we proceed through the following steps.

\begin{enumerate}
	\item Compute $w_{max}=\max_{x \in \mathcal{R}_i} w_{i}(x)$.
	\item Generate $x_{prop}$ over $\mathcal{R}_i$ from a homogeneous Poisson process with intensity $w_{max}$.
	\item Compute $p(x_{prop}) = w_{i}(x_{prop})/w_{max}$.
	\item Generate a sample $u$ from the uniform distribution on $(0,1)$.
	\item If $k=1$, set $x_{1}=x_{prop}$ if $u \leq p(x_{prop})$; for $k>1$ and given $\{x_{j}: j=1,\ldots,k-1\}$,  set $x_{k}=x_{prop}$ if $u \leq p(x_{prop})$ and $x_{prop}$ falls at the intersection of $\mathcal{R}_i$ and $\{x \in \mathcal{R}_i: \|x-x_{j}\| > \delta (1-w(x_{j})/w_{max})\}$. Otherwise, reject $x_{prop}$.
	\item Repeat 2 to 5, until $k=L_{i}$.
\end{enumerate}

To identify a suitable value for $L_{i}$ (the total number of generated points within $\mathcal{R}_i$), a possible solution is to use the packing density for a sequential inhibitory point process given by 
\begin{equation}
\label{eq:packing_dist}
\gamma = \frac{L_i \pi \delta^2}{4 |\mathcal{R}_i|},
\end{equation}
where $\delta$ is the minimum permissible distance between points. The maximum possible value for $\gamma$ is obtained by close-packed discs whose centres form an equilateral triangular lattice with sides of length $\delta =\pi/\sqrt{12}$.  Through a simulation study, Tanemura \cite{tanemura1979random} suggested to set $\gamma=0.55$ in order to achieve good spatial coverage in a relatively small number of iterations. Once $\gamma$ and $\delta$ are fixed, we can then obtain $L_i$ through equation \eqref{eq:packing_dist}.

An alternative solution is to leave choose $\gamma$ as a function of $\phi$ using the following adaptive algorithm.

\begin{enumerate}
	\item For a given $\phi$, initialize a batch size $k$ and a relative tolerance $\epsilon$;
	\item Locate $k$ quadrature points with packing intensity $\gamma(k)=k \pi \delta^2/4 |\mathcal{R}_i|$, evaluate the integral in \eqref{eq:IntegralApprox} and denote its value as $I_{old}$;
	\item Add $k$ points using a packing intensity $\gamma(k)/2$, re-evaluate the integral and denote its value as $I_{new}$;
	\item If $I_{new} = I_{old}$, stop the algorithm. Otherwise, set $I_{new}=I_{old}$, add $k$ points with $\gamma(k)/3$ and repeat until $|I_{old} - I_{new}| < \epsilon |I_{new}|.$	
\end{enumerate}

Since $\phi$ is almost always unknown, the adaptive algorithm becomes more computationally demanding, especially in the case of a large number of regions in the study domain and for small values of $\phi$ which require a finer grid for a satisfactory approximation of \eqref{eq:covariance}. When fitting the model in \eqref{eq:approx} (see next section for more details), our recommendation is to use the non-adaptive algorithm first, in order to locate the likely value of $\phi$, and then to run a final estimation using the adaptive algorithm. In the application in Section \ref{sec:application}, the adaptive algorithm increases the elapsed time by about 10 minutes (592 seconds) on a laptop with 7.6GiB memory and $2.40\text{GHz} \times 4$ processor. Furthermore, in order to reduce the computational burden, we propose to discretise $\phi$ over a finite set of values and pre-compute the covariance matrix as defined by \eqref{eq:IntegralApprox} for each of the pre-defined values. To obtain a 95$\%$ confidence interval for $\phi$, we then compute the profile likelihood over the discrete set and interpolate it using a natural cubic spline. In our experience, the fineness of the discretisation does not have tangible effects on the spatial predictions but, instead, directly affects the goodness of the numerical approximation of the $95\%$ confidence interval based on the profile likelihood.  \par

\subsection{Monte Carlo maximum likelihood}
\label{subsec:mcml}
We carry out parameter estimation for the model in \eqref{eq:approx} using the Monte Carlo maximum likelihood (MCML) method.\citep{christensen} \par

Let $f(\cdot)$ be a shorthand notation for ``the density function of $\cdot$''.
Let $y^\top=(y_1,\ldots,y_n)$ and linear predictor $\eta^\top=(\eta_{1},\ldots,\eta_{n})$; it then follows that conditionally on $S^*=(S_1^*,\ldots,S^*_n)^\top$, the joint distribution of $Y$ is 
\begin{equation*}
f(y|\eta) = \prod_{i=1}^n f(y_i|\eta_i),
\end{equation*}
where
$$
f(y_i|\eta_i) \propto \exp\{y_i \log \mu_i-\mu_i\}.
$$
Let $\psi= (\beta, \sigma^2,\phi)$ denote the vector of the model parameters, then the likelihood function for $\psi$ is obtained by integrating out $S^*$, i.e.
\begin{eqnarray} 
\label{eq:lik1}
L(\psi) &=& \int_{\mathbb{R}^n} f(y | \eta)~f(\eta; \psi) \: d\eta. ~~~~~~~~~~~~~~~~~
\end{eqnarray}
In \eqref{eq:lik1} $f(\eta; \psi)$ is a multivariate Gaussian distribution function with mean $D \beta$, where $D$ denotes a matrix of explanatory variables, and covariance matrix $\Sigma$, whose $(i,j)$-th entry is given by \eqref{eq:covariance}. To reduce the computational burden accrued from the numerical approximation \eqref{eq:IntegralApprox}, we restrict the maximization of \eqref{eq:lik1} to a finite set of predefined values for $\phi$ and, for each of these, pre-compute the covariance matrix $\Sigma$ together with its inverse, determinant and Cholesky decomposition. \par

Since the high-dimensional integral in \eqref{eq:lik1} cannot be solved analytically, we use Monte Carlo methods for the approximation of the likelihood. Let $\psi_0$ denote our best guess of $\psi$. We re-write \eqref{eq:lik1} as
\begin{eqnarray} 
\label{eq:lik2}
L(\psi) &\propto &  E_{\eta | y} \left[\frac{f(\eta; \psi)}{f(\eta; \psi_0)} \right],
\end{eqnarray}
where the expectation $E$ is taken with respect to the conditional distribution of $\eta$ given $y$ with parameters vector $\psi_{0}$. We provide the proof of this in Appendix A of the supplementary material. \par

To generate $N$ samples, say $\eta_{(j)}$, from the conditional distribution of $\eta$ given $y$, we use a Monte Carlo Markov chain (MCMC) algorithm implemented in the \texttt{Laplace.sampling.MCML} function in the PrevMap package.\citep{giorgi2017} This function uses a Metropolis-adjusted Langevin MCMC algorithm to update the standardised vector of random effects, $\tilde{\eta}=\hat{\Sigma}^{-\tfrac{1}{2}} (\eta-\hat{\eta})$, where $\hat{\eta}$ and $\hat{\Sigma}$ are the mode and the inverse of the negative Hessian of  $f(\eta; \psi_0)$ at $\hat{\eta}$. We can then approximate the likelihood function in \eqref{eq:lik2} as
\begin{eqnarray} 
\label{eq:lik_approx}
L(\psi) \approx L_N(\psi) &=& \frac{1}{N}~ \sum_{j=1}^N~\frac{f(\eta_{(j)}; \psi)}{f(\eta_{(j)}; \psi_0)}.
\end{eqnarray}
As $N \rightarrow \infty$, in the above equation, $L_{N}(\psi)$ converges to $L(\psi)$. \cite{geyer1992, geyer1994, geyer1996}

Finally, we maximize \eqref{eq:lik_approx} using a  constrained quasi-Newton optimization algorithm, implemented in the \texttt{nlminb} function in the R software environment, by providing analytical expressions for the first and second derivatives of \eqref{eq:lik_approx} with respect to $\psi$. If $\hat{\psi}_N$ denote the resulting MCML estimate, we then set $\psi_0=\hat{\psi}_N$ and repeat the previous steps until convergence.

\subsection{Continuous spatial prediction}
\label{subsec:predictions}

We now consider the problem of carrying out spatial prediction of $S^*(x)$ at a pre-defined location $x$ within the study area $A$. Using the same notation as in the previous section, we first note that
\begin{eqnarray}
\label{eq:spat_pred}
f(S^*(x) | y) &=& \int_{\mathbb{R}^n} f(\eta, S^*(x) | y) \: d \eta \nonumber \\
&=& \int_{\mathbb{R}^n} f(\eta | y) f(S^*(x) | \eta, y)\: d \eta \nonumber \\
&=& \int_{\mathbb{R}^n} f(\eta | y) f(S^*(x) | \eta) \: d \eta. 
\end{eqnarray}
Hence, we sample from $f(S^*(x) | y)$ using the following two-step procedure: (1) draw samples $\eta_{(j)}$, for $j=1,\ldots,N$ from $f(\eta | y)$ using the MCMC algorithm described in the previous section; (2) for each $\eta_{(j)}$, for $j=1,\ldots,N$  simulate from $f(S^*(x) | \eta_{(j)})$,  a Gaussian distribution with mean 
$
\mu^*(x) = c(x)^\top\Sigma^{-1}(\eta_{(j)}-D\beta)
$
and variance
$
v^2(x) = \sigma^2-c(x)^\top \Sigma^{-1} c(x),
$
where $c(x)^\top = (c_{1}(x), \ldots, c_{n}(x)),$ 
$
c_{i}(x) = \sigma^2 \int_{\mathcal{R}_{i}} w(x) \rho(\|x-x'\|) \: dx' 
$, and we use \eqref{eq:IntegralApprox} to approximate the integral. The resulting samples from $f(\eta | y)$ can then be used to compute non-linear properties of $S^*(x)$ and to summarise these using, for example, predictive means and standard errors.

\section{Simulation Study} 
\label{sec:simulation}
We now conduct a simulation study to assess the predictive performance of the proposed approximation in \eqref{sec:sdlgcp} when the underlying process is an LGCP model. \par
We simulate $B=1,000$ data-set of counts using the administrative boundaries of the lower layer super output areas (LSOAs) in Newcastle-Upon-Tyne, UK, as in the application of Section \ref{sec:application}. We specify the offsets $m(x)$ using population density estimates from the OpenPopGrid database \citep{murdock2015openpopgrid} and  simulate the locations of the events using an inhomogeneous Poisson process with intensity $m(x) \exp\{S(x)\}$. We define three scenarios by setting the standard deviation of the Gaussian random field $S(x)$ to $\sigma=0.706$ and let $\phi$ (whose unit of measure is metres) vary over the set $\{100,800,1500\}$, which correspond to a case of small, medium and large spatial correlation, respectively. The value of the standard deviation corresponds to the posterior mean obtained from the fitted LGCP in the application to primary biliary cirrhosis data described in the next section. Finally, for each of the $1,000$ simulated data-sets of aggregated counts at LSOA level, we fit the following models.

\begin{itemize}	
	\item \textit{LGCP.} We use a Bayesian data augmentation technique implemented in the \texttt{lgcp} package.\citep{taylor2015bayesian} We overlay a computational grid at a spacing of of $300 \times 300$ metres onto the area of interest and fit the model in \eqref{eq:lgcpmodel}.  We run 3,100,000 iterations of the MCMC algorithm with a burn-in of 100,000 samples and then retain every 300-th sample.
	
	\item \textit{Spatially discrete approximation (SDA) to LGCP.} We fit the approximation in \eqref{sec:sdlgcp} using a population weighted average (SDA I, with $w_{i}(x)=m(x)/m_i$) and simple average (SDA II, with $w_{i}(x)=1/|\mathcal{R}_i|$) of the log-intensity. For both, we use the MCML method described in Section \ref{subsec:mcml} and run 110,000 iterations of the MCMC algorithm with a burn-in of 10,000 samples and then retain every 10-th sample. 
	
\end{itemize}

We summarise the results from the simulation  study through the bias, root-mean-square-error (RMSE), width of the predictive interval (WPI) and the 95$\%$ coverage probability (CP) for the incidence at LSOA level, $\lambda_i$, and for the spatially continuous relative risk, $\exp\{S(x)\}$. Let $\lambda_{i}^{(j)}$ denote the true simulated incidence for $\mathcal{R}_i$ at the $j$-th simulation; hence
\begin{equation*}
BIAS = \frac{1}{n B} \sum_{i=1}^{n} \sum_{j = 1}^{B} (\hat{\lambda}_{i}^{(j)} - \lambda_{i}^{(j)}),
\end{equation*}
\begin{equation*}
RMSE =  \sqrt{\frac{1}{n B} \sum_{i=1}^{n} \sum_{j = 1}^{B} (\hat{\lambda}_{i}^{(j)} - \lambda_{i}^{(j)})^2},
\end{equation*}
\begin{equation*}
WPI = \frac{1}{n B} \sum_{i=1}^{n} \sum_{j = 1}^{B} (PI_{0.95,U}^{(j)}-PI_{0.95,L}^{(j)}),
\end{equation*}
\begin{equation*}
CP = \frac{1}{n B} \sum_{i=1}^{n} \sum_{j = 1}^{B} I(\lambda_{i}^{(j)} \in PI_{0.95}^{(j)}),
\end{equation*}
where $\hat{\lambda}_{i}^{(j)}$ is the mean of the predictive distribution for $\lambda_{i}^{(j)}$, $I(\lambda_{i}^{(j)} \in PI_{0.95}^{(j)})$ is an indicator function that takes value 1 if $\lambda_{i}^{(j)}$ falls inside the 95$\%$ prediction interval and 0 otherwise, and $PI_{0.95,U}^{(j)}$ and $PI_{0.95,L}^{(j)}$ are the upper and lower limits of the 95$\%$ prediction interval, respectively. Similarly, we compute the three indices for the relative risk $\exp\{S(x)\}$ by averaging each of these over the regular grid at a spacing of $300$ metres covering the whole of Newcastle-Upon-Tyne, UK.

Table \ref{table:simulation:one} reports the results for the prediction of $\lambda_i$, the incidence at LSOA level. We observe that SDA I and II have a slightly lower bias and RMSE than LGCP in all three scenarios, with SDA I having the best performance. The coverage probability is close to the $95\%$ nominal level and the WPI is comparable for all three models. \par

The results for the spatially continuous relative risk, $\exp\{S(x)\}$, are shown in Table \ref{table:simulation:two}. LGCP has the lowest bias and RMSE followed by SDA I in all three scenarios, with larger differences for $\phi=800$ and $\phi=1500$. Both SDA I and II are more conservative than LGCP and provide prediction intervals with a larger coverage than the nominal level, as the result of a large RMSE. We also observe that the use of the population weighted average in SDA I leads to a tangible reduction in RMSE and bias with respect to SDA II.

\section{Application: mapping of primary biliary cirrhosis risk} 
\label{sec:application}
We analyse incidence data on primary biliary cirrhosis (PBC) in Newcastle-Upon-Tyne, UK, obtained from the original study carried out by Princ et al.\cite{prince2001geographical}; the data-set is freely available from the \texttt{lgcp} R package. The data consist of geo-referenced cases of definite or probable PBC between 1987 and 1994. The objective of this analysis is to quantify the difference in the predictive inferences between the gold-standard LGCP model and the proposed spatially discrete approximation (or SDA), on PBC incidence at LSOA level and the spatially continuous relative risk surface. In the case of SDA, we fit the population weighted (SDA I) and simple average (SDA II) versions described in the previous section. We also consider the exponential variogram (EV) model proposed by Wall\cite{wall2004close} consisting of a geostatistical Poisson model for the counts whose spatial structure is defined using the centroids of each LSOA. Finally, we fit the Besag, York and Molli\'{e} \cite{besag1991bayesian} (BYM) model, one of most commonly used approaches in small area estimation, with linear predictor
$$
\log{\lambda_{i}} = d_{i}^\top \beta^* + S_{i} + Z_{i}
$$
where $S_{i}$ is a zero-mean intrinsically autoregressive process with variance $\sigma^2$ and $Z_{i}$ is Gaussian noise with variance $\tau^2$. 

In all five models, we use the index of multiple deprivation (IMD) as a covariate of the linear predictor. The IMD is publicly available from the UK Government online archives (\url{webarchive.nationalarchives.gov.uk}). The regression coefficients for the IMD are denoted by $\beta_{i}$ in the LGCP model and by $\beta_{i}^*$ in the BYM, EV and SDA models, with $i=0$ corresponding to the intercept and $i=1$ the effect of IMD. \par

For the SDA models, we run 110,000 iterations of the MCMC algorithm with a burn-in of 10,000 samples, and then retain every 10-th sample. We discretise $\phi$ using 100 equally spaced values between 50 and 2000 meters. \par

For the LGCP model, we specify independent priors as follows: $\log \sigma \sim N(\log 1, 0.15)$, $\log \phi \sim N(\log 500, 2)$ and $(\beta_{0},\ldots,\beta_{7}) \sim MVN(0, 10^6I)$. We run 3,100,000 iterations of the MCMC algorithm with a burn-in of 100,000 samples and retain every 3000-th sample so as to obtain a set of 1,000 weakly dependent samples.  \par

Fitting of the BYM model using \texttt{CARBayes}\cite{carbayes} is carried out by iterating the MCMC algorithm 1,100,000 times with a burn-in of 100,000 samples and retaining every 100-th sample. \par

Finally, for the EV model which fit using the \texttt{spBayes} \cite{spbayes} R package, we specify an inverse-Gamma prior on the variance parameter $\sigma^2$ with shape parameter 1 and scale parameter 2. The spatial scale parameter $\phi$ is assigned a uniform prior in the interval $[50, 2500]$. For the regression coefficients $\beta$, we use a flat prior. We run 1,100,000 iterations of the MCMC algorithm with a burn-in of 100,000 samples and retain every 40-th sample. \par

Trace-plots and correlograms are used to assess convergence of the MCMC algorithms in each of the fitted models. These are reported in the Appendix, from section B to E, and all indicate a good mixing of the resulting MCMC samples. \par

Tables \ref{table:three} reports the point and interval estimates for the parameters of each of the fitted models. We observe that the differences amongst the point estimates of the regression coefficients from the five models are small. \par

\begin{figure}[htbp]
	\centering
	\fbox{\begin{minipage}{12 cm}
			% add text or graphics here
			\centering
			\begin{adjustbox}{scale=1,clip,trim=0cm 0cm 0cm 0cm}
				\includegraphics[scale = 0.5]{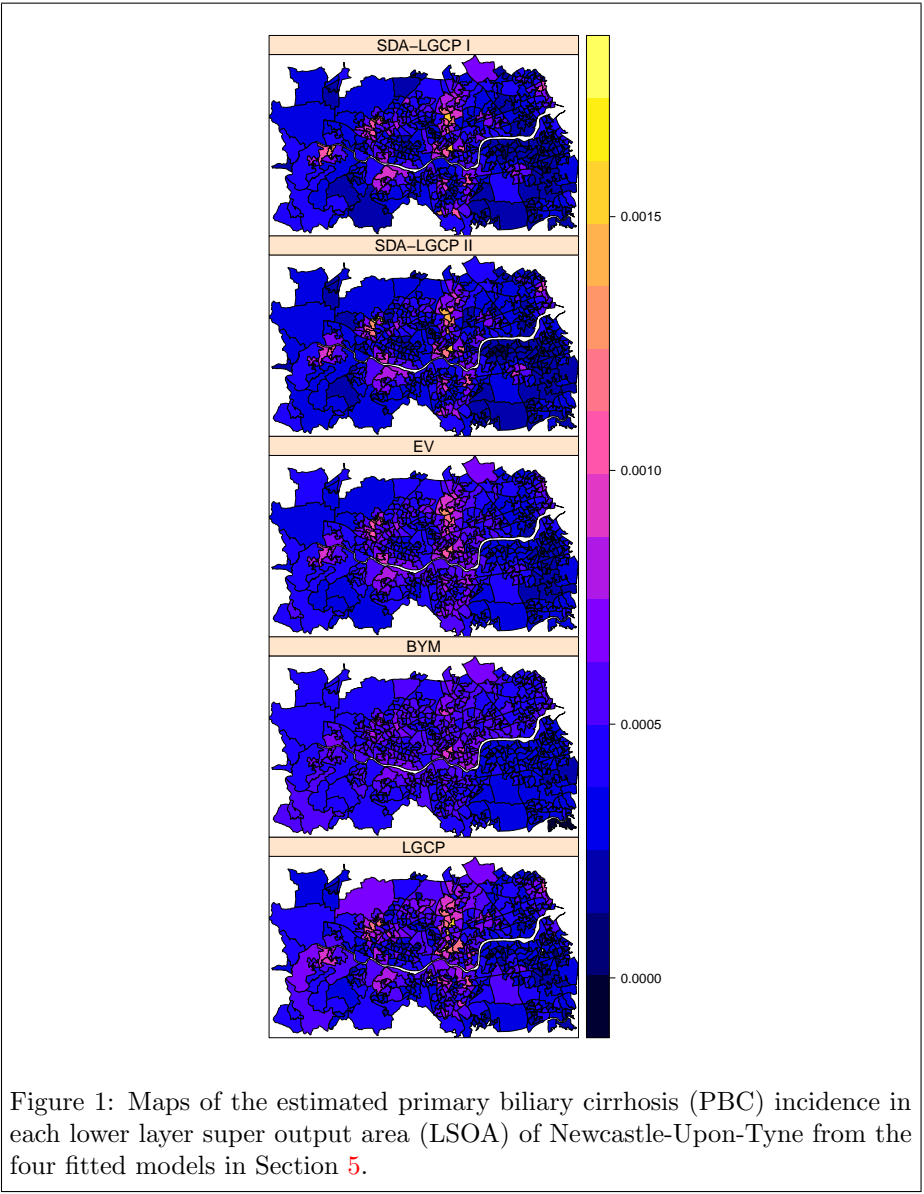} 
			\end{adjustbox}
			\caption{Maps of the estimated primary biliary cirrhosis (PBC) incidence in each lower layer super output area (LSOA) of Newcastle-Upon-Tyne from the four fitted models in Section \ref{sec:application}.}
			\label{fig:one}
	\end{minipage}}
\end{figure}

\begin{figure}[htbp] 
	\centering
	\fbox{\begin{minipage}{12 cm}
			% add text or graphics here
			\centering
			\includegraphics[scale = 0.5]{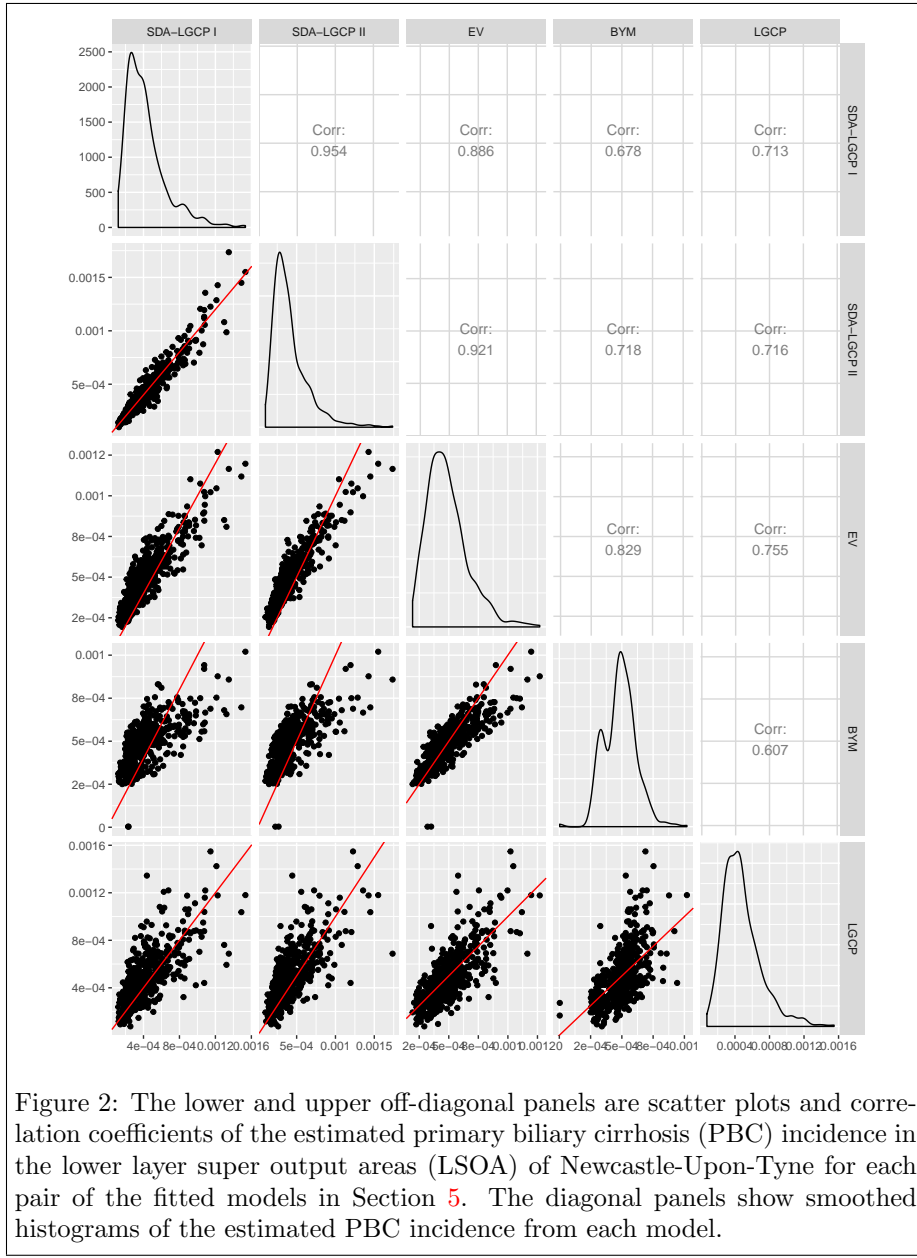} 
			\caption{The lower and upper off-diagonal panels are scatter plots and correlation coefficients of the estimated primary biliary cirrhosis (PBC) incidence in the lower layer super output areas (LSOA) of Newcastle-Upon-Tyne for each pair of the fitted models in Section \ref{sec:application}. The diagonal panels show smoothed histograms of the estimated PBC incidence from each model.}
			\label{fig:two}
	\end{minipage}}
\end{figure}

\begin{figure}[htbp] 
	\centering
	\fbox{\begin{minipage}{12 cm}
			% add text or graphics here
			\centering
			\includegraphics[scale = 0.5]{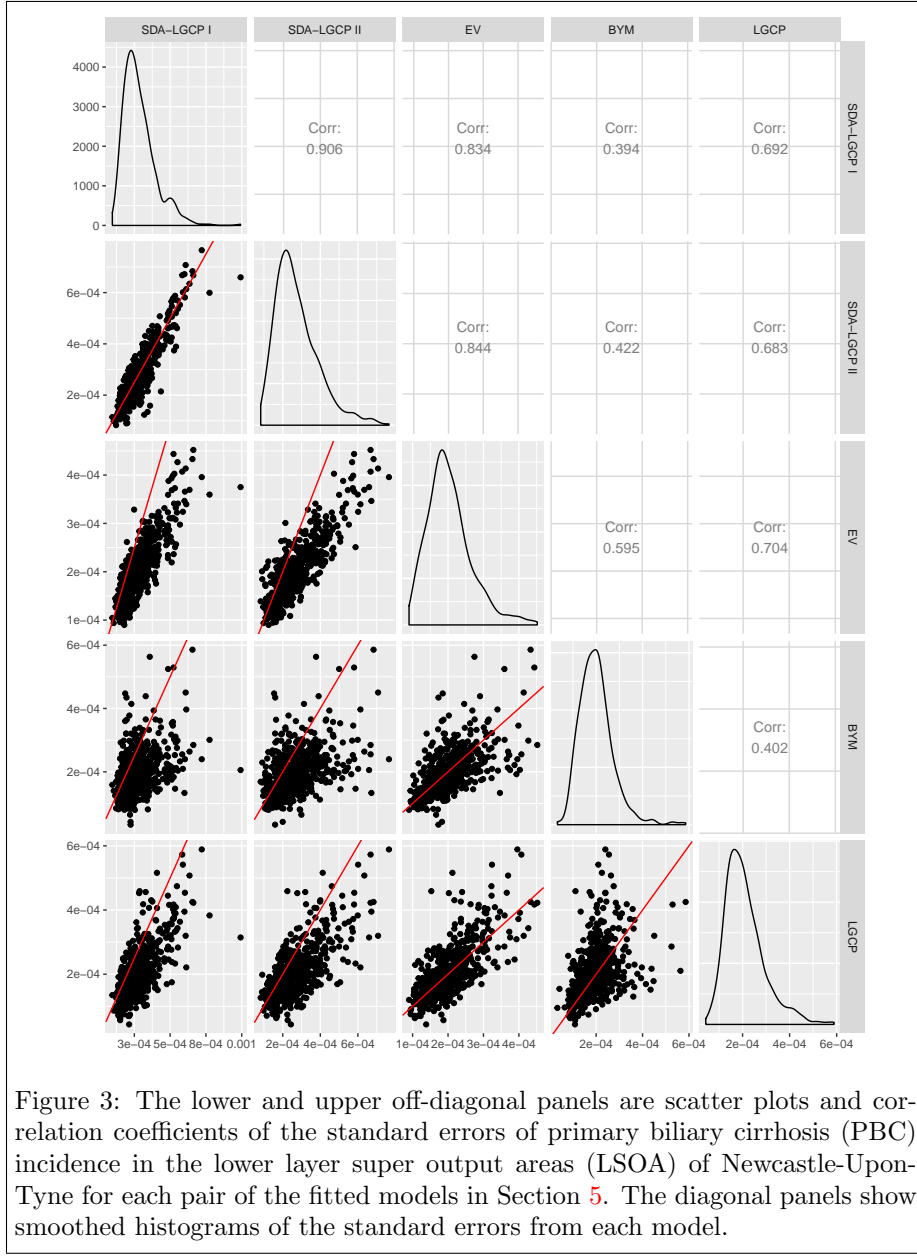} 
			\caption{The lower and upper off-diagonal panels are scatter plots and correlation coefficients of the standard errors of primary biliary cirrhosis (PBC) incidence in the lower layer super output areas (LSOA) of Newcastle-Upon-Tyne for each pair of the fitted models in Section \ref{sec:application}. The diagonal panels show smoothed histograms of the standard errors from each model.}
			\label{fig:three}
	\end{minipage}}
\end{figure}

Figure \ref{fig:one} shows a map of the estimated PBC incidence at LSOA level from the five models. The spatial spatial pattern estimated by each of these is comparable, as indicated by the scatter plots of Figure \ref{fig:two}. The same consideration holds for the predictive standard errors  (Figure \ref{fig:three}). More specifically, the estimated incidence from the LGCP model has a correlation of about 0.7 with the other models, expect the BYM model for which the correlation is about 0.6. The good performance of the EV model can be explained by the fact that, in this scenario, the size of most of the LSOAs is small relative to the range of the spatial correlation, hence the use of the centroid becomes less problematic. \par

\begin{figure}[htbp]
	\centering
	\fbox{\begin{minipage}{12cm}
			% add text or graphics here
			\centering
			
			\begin{adjustbox}{scale=0.5,clip,trim=0cm 0cm 0cm 0cm}
				\includegraphics{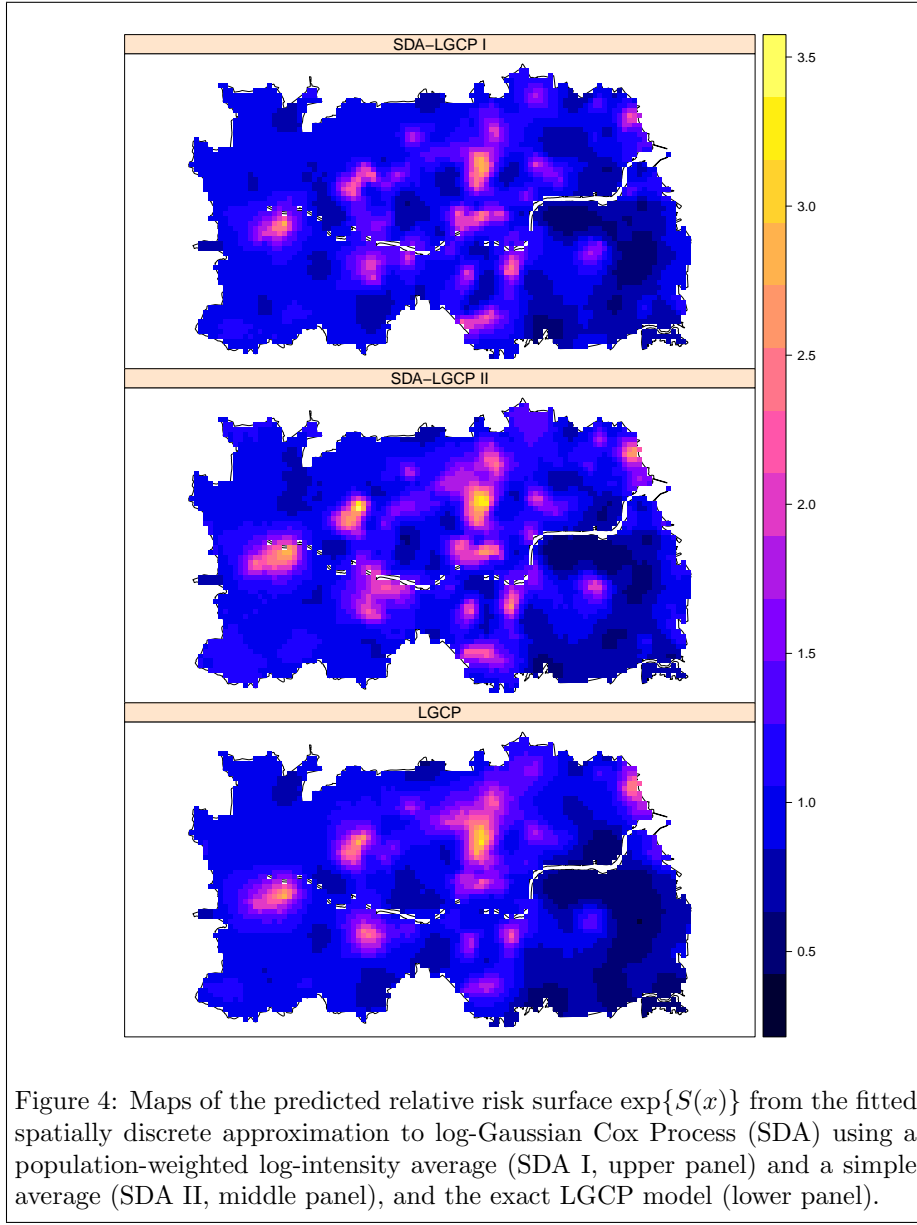} 
				% ...sweave content
			\end{adjustbox}
			\caption{Maps of the predicted relative risk surface $\exp\{S(x)\}$ from the fitted spatially discrete approximation to log-Gaussian Cox Process (SDA) using a population-weighted log-intensity average (SDA I, upper panel) and a simple average (SDA II, middle panel), and the exact LGCP model (lower panel). }
			\label{fig:four}
	\end{minipage}}
\end{figure}

\begin{figure}[htbp]
	\centering
	\fbox{\begin{minipage}{12cm}
			% add text or graphics here
			\centering
			\includegraphics[scale = 0.5]{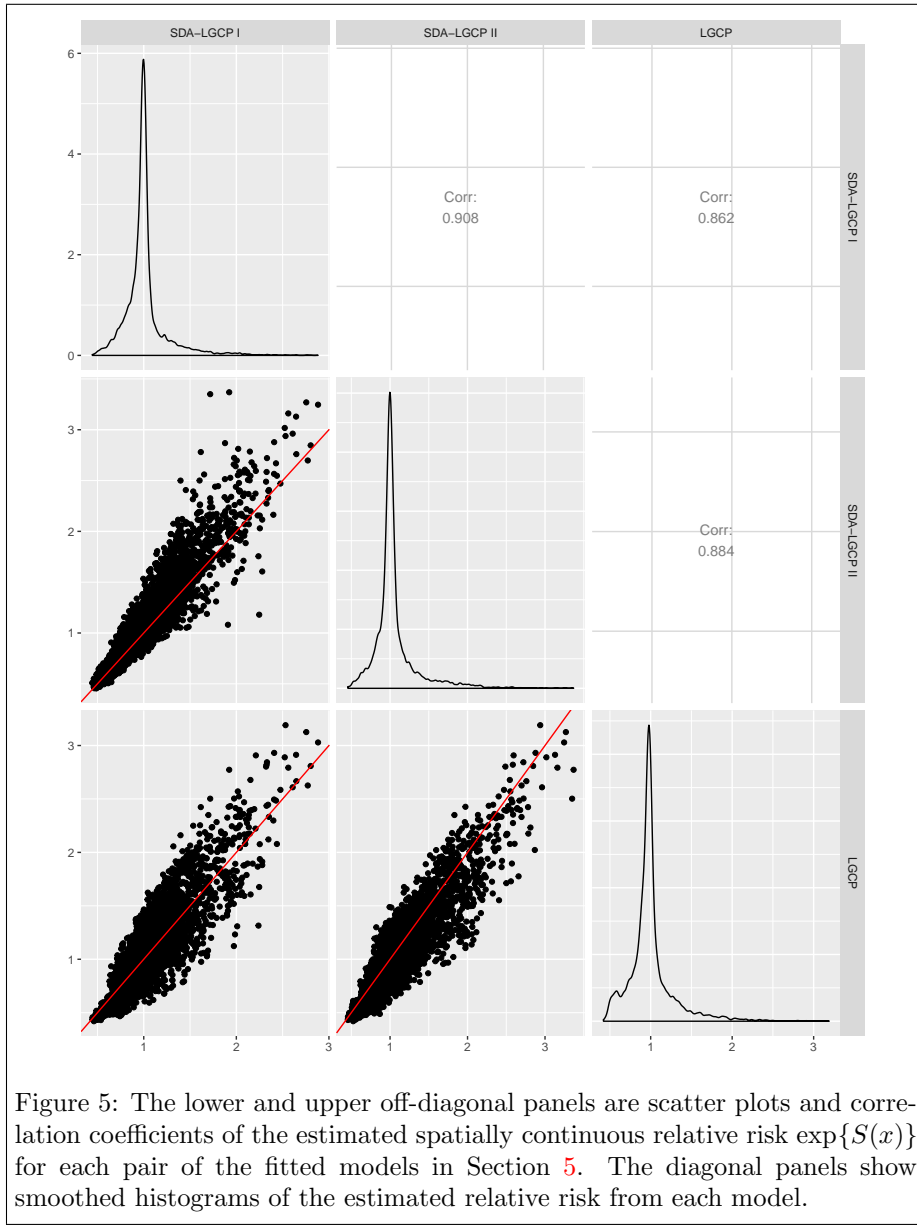} 
			\caption{The lower and upper off-diagonal panels are scatter plots and correlation coefficients of the estimated spatially continuous relative risk $\exp\{S(x)\}$ for each pair of the fitted models in Section \ref{sec:application}. The diagonal panels show smoothed histograms of the estimated relative risk from each model.}
			\label{fig:five}
	\end{minipage}}
\end{figure}

\begin{figure}[htbp]
	\centering
	\fbox{\begin{minipage}{12cm}
			% add text or graphics here
			\centering
			\includegraphics[scale = 0.5]{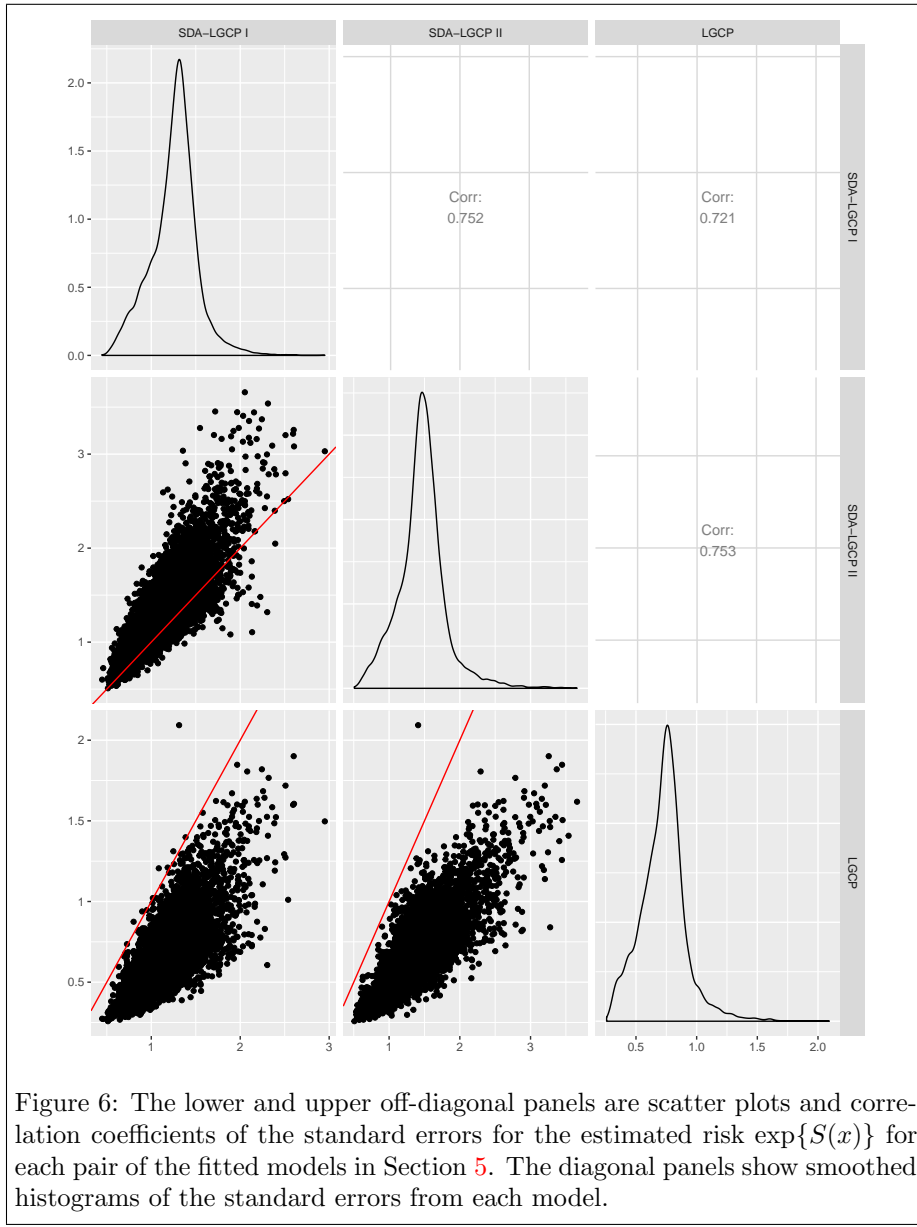} 
			\caption{The lower and upper off-diagonal panels are scatter plots and correlation coefficients of the standard errors for the estimated risk $\exp\{S(x)\}$ for each pair of the fitted models in Section \ref{sec:application}. The diagonal panels show smoothed histograms of the standard errors from each model.}
			\label{fig:six}
	\end{minipage}}
\end{figure}

Figure \ref{fig:four} shows the map of the estimated continuous relative risk surface $\exp\{S(x)\}$ over a $300 \times 300$ meters regular grid covering the whole of the study area. The scatter plots (Figures \ref{fig:five} and \ref{fig:six}) indicate that the point estimates from the LGCP and the SDA approach are strongly similar, with a correlation of 0.862 between SDA I and LGCP and of 0.884 between SDA II and LGCP. However, we also observe that the standard errors from SDA, both I and II, are larger than those from LGCP. This is consistent with our results from the simulation study of the previous section.

%\begin{figure}[htbp]
%	\centering
%	\fbox{\begin{minipage}{12cm}
%			% add text or graphics here
%			\centering
%			\includegraphics[scale = 0.5]{IMAGES/corrMatrix.pdf} 
%			\caption{Images of the estimated correlation matrices for the spatially discrete approximation to the log-Gaussian Cox process with a population weighted district-wide average, denoted by SDA I (upper panel), SDA II using a simple district-wide average (middle panel) the conditional autoregressive model (lower panel).}
%			\label{fig:corr}
%\end{minipage}}
%\end{figure}

\section{Discussion} 
\label{sec6}
In this article we have developed a spatially discrete approximation (SDA) to log-Gaussian Cox process (LGCP) models in order to carry out spatial prediction of disease risk at any desired spatial scale using spatially aggregated disease count data. \par 

As variation in disease risk occurs in a spatial continuum irrespective of the format in which the data are available, we consider the LGCP framework to be a natural statistical paradigm for modelling aggregated disease count data. However, when computational constraints make the fitting of an LGCP infeasible, we argue that SDA provides a computationally efficient solution while respecting the spatially continuous nature of disease risk. SDA also overcomes some of the limitations inherent to other spatially discrete models, such as CAR models. In addition to providing spatially continuous predictions, SDAs can also deal with the issue of changing administrative boundaries over time and allow incorporation of covariates available at any spatial scale. 

Kelsall et al \cite{kelsall2002modeling} developed a similar approach to the proposed SDA for modelling count data available at areal level. Specifically, by assuming an intercept-only model, they approximate \eqref{eq:lgcpmodel} using a multivariate log-Gaussian distribution with mean 
$$
E[\lambda_{i}] = \exp\{\beta_0+\sigma^2/2\}
$$
and covariance
\begin{eqnarray*}
	{\rm Cov}\{\lambda_{i}, \lambda_{j}\} &=& \exp\{\beta_0+\sigma^2/2\} \times \\ &&\left[\int_{\mathcal{R}_i}\int_{\mathcal{R}_j} w_{i}(x)w_{j}(x')\exp\{\sigma^2 \rho(\|x-x'\|;\phi)\} \: dx \: dx' -1\right].
\end{eqnarray*}
Kelsall et al\cite{kelsall2002modeling} then advocate the use of the log-Gaussian approximation as a Bayesian prior for spatial smoothing but no reference is made to the LGCP framework. In this paper, instead, our objective was to develop a computationally efficient approximation to the LGCP model which, in Bayesian terms, is our chosen prior for modelling disease risk. \par

In fitting SDA models, most of the computational burden is due to the approximation of the integral in \eqref{eq:covariance}, which defines the area-level correlation between the spatial random effects. In our example, the SDA model is about 5 to 15 times faster to fit than the LGCP model, depending on the number of values used to discretise the scale of the spatial correlation $\phi$. To make SDA even faster, efficient approximations to Gaussian processes should also be considered (see, for example, Lindgren et al\cite{lindgren2011explicit}). These could be used to sample from the predictive distribution of $S^*(x)$ in \eqref{eq:log_intens} and avoid computation of the integral in \eqref{eq:covariance}. \par

We conclude that SDA is a reliable approximation to LGCP for carrying out predictions at areal-level, both in terms of point predictions and in the quantification of uncertainty. It also provides spatially continuous predictions in disease risk that are comparable to those from LGCP, but with larger standard errors and more conservative predictions intervals. \par

Finally, extension to the spatio-temporal case of the method discussed in this paper is possible and is work in progress. For example, let us consider counts $y_{it}$ for the region $\mathcal{R}_i$ over the time interval $(t, t+1)$. Let $S(x,t)$ be a spatio-temporal Gaussian process with covariance function 
$$
{\rm cov}\{S(x,t), S(x',t')\} = \sigma^2 \exp\{-|t-t'|/\psi\}\exp\{-\|x-x'\|/\phi\}.
$$ By modelling the $y_{it}$ as realisations of a spatio-temporal log-Gaussian Cox process with conditional intensity $\Lambda(x,t)=\exp\{\alpha+S(x,t)\}$, 
we can then approximate this with a spatio-temporally discrete Gaussian process $S^*_{t}=(S^*_{1t}, \ldots, S^*_{nt})$, such that
$$
S^*_{t} = \varphi S_{t-1}^* + W_{t}, 0 < \varphi < 1,
$$
where the temporal innovation $W_{t}$ is modelled as a multivariate Gaussian distribution with covariance matrix given by \eqref{eq:covariance}.
Preliminary results suggest that the reduction in computing time with respect to a spatio-temporal LGCP model is substantially larger than that observed for the purely spatial scenario presented in this paper.

\section*{Acknowledgements}
We thank Dr Benjamin Taylor for his useful comment on the draft of this manuscript. We also acknowledge the travel grant from African Institute for Mathematical Sciences to OOJ.

\section*{Funding}
OOJ holds a Connected Health Cities funded PhD studentship.

%%%%%%%%%%%%%%%%%%%%%%%%%%% map of the simulation

%%%%%%%%%%%%%%


%\begin{sm}
%To typeset a
%"Supplemental material" section.
%\end{sm}

%\bibliographystyle{plain}
%Harvard (name/date)
%\bibliographystyle{WileyNJD-AMA}
\bibliographystyle{apalike}
\bibliography{Paper1}
%%Vancouver (numbered)
%\bibliographystyle{SageV} 
%\bibliographystyle{apalike}
\newpage
\begin{table}[htp]
	\begin{center}
		\caption{Average bias, root-mean-square-error (RMSE), width of the $95\%$ prediction interval (WPI) and the 95\% coverage probability (CP) for the LSOA incidence, $\lambda_i$, from the simulation study of Section \ref{sec:simulation}.} \label{table:simulation:one}
		\scalebox{0.9}{
			\begin{tabular}{lccc|ccc|ccc}
				\hline
				\multicolumn{4}{l|}{ \quad\quad  \quad \quad \quad \quad \quad \quad $\phi = 100$}  & \multicolumn{3}{l|}{\quad\quad  \quad \quad $\phi = 800$}  & \multicolumn{3}{l}{\quad\quad  \quad \quad $\phi = 1500$}\\
				
				&	SDA I & SDA II & LGCP & 
				SDA I & SDA II & LGCP &    
				SDA I & SDA II & LGCP 
				\\
				\hline
				Bias  & -0.006 & -0.007 & -0.009 & -0.002 & -0.003 & -0.004 &  -0.008 & -0.008 &  -0.011 \\
				RMSE  & 0.020 & 0.021 & 0.022 & 0.003 & 0.004 & 0.006 &  0.027 &  0.029 & 0.030 \\
				WPI & 0.015 &   0.016 & 0.017 &  0.002 & 0.003 & 0.004 &  0.026 & 0.028 & 0.028 \\
				95\%CP & 0.940 &   0.942 & 0.948 &  0.942 & 0.943 & 0.952 &  0.943 & 0.944 & 0.945 \\
				\hline
		\end{tabular}}
	\end{center}
\end{table}

\newpage
\begin{table}[htp]
	\begin{center}
		\caption{Average bias, root-mean-square-error (RMSE), width of the 95$\%$ prediction interval (WPI) and the 95\% coverage probability (CP) for the spatially continuous relative risk, $\exp\{S(x)\}$, from the simulation study of Section \ref{sec:simulation}. \label{table:simulation:two}}
		\scalebox{0.9}{
			\begin{tabular}{lccc|ccc|ccc}
				\hline
				\multicolumn{4}{l|}{ \quad\quad  \quad \quad \quad \quad \quad \quad $\phi = 100$}  & \multicolumn{3}{l|}{\quad\quad  \quad \quad $\phi = 825$}  & \multicolumn{3}{l}{\quad\quad  \quad \quad $\phi = 1500$}\\
				&	SDA I & SDA II & LGCP &
				SDA I & SDA II & LGCP &     
				SDA I & SDA II & LGCP \\
				\hline
				Bias  & -0.575  & -0.582 & -0.566 &  0.842 & 0.965 & -0.108 &  0.299 &  0.316  &  0.227 \\
				RMSE  & 2.590 & 2.800 & 0.045 & 0.439 & 0.531 & 0.005 &  2.070 &  2.260  & 0.002 \\
				WPI & 2.525 &   2.739 & 0.564 &  0.719 & 0.806 & 0.108 &  2.048 & 2.238 & 0.227 \\
				95\%CP & 0.988 &   0.990 & 0.940 &  0.979 & 0.983 & 0.948 &  0.975 & 0.982 & 0.942 \\
				\hline
		\end{tabular}}
	\end{center}
\end{table}

%%%%%%%%%%%%%%%%%%%%%%%%%%%%%%%%%%%%%%%
\newpage
\begin{table}[htbp]
	\begin{center}
		\caption{Point estimates and 95$\%$ confidence/credible intervals (CI) for the model parameters of the spatially discrete approximation to log-Gaussian Cox Process (LGCP) using a population-weighted log-intensity average (SDA I) and a simple average (SDA II), the exponential variogram (EV) model, Besag-York-Molli\'e (BYM) model and the LGCP model.} \label{table:four}
		\label{table:three}
		\begin{tabular}{clcc}
			\hline
			\bf{Model} & \bf{Parameter} & \bf{Estimate} & \bf{95$\%$ CI} \\	
			\hline
			SDA I & $\sigma^2$ &  1.043 & (0.907,  1.180) \\
			&			$\phi$ & 742.857 & (453.153, 1005.405)\\
			&			$\beta_{0}^*$ & -8.080 & (-8.248, -7.912)\\
			&			$\beta_{1}^*$ & 0.008 & (0.004,  0.011)\\
			\hline 
			SDA II & $\sigma^2$ &  1.020 & (0.898,  1.142) \\
			&	$\phi$ & 857.143 & (489.590 1037.638)\\
			&		$\beta_{0}^*$ & -7.876 & (-8.029, -7.722)\\
			&			$\beta_{1}^*$ & 0.006 & (0.002,  0.010)\\
			\hline 
			EV & $\sigma^2$ &  0.316 & (0.246, 0.369) \\
			&			$\phi$ & 525.570 & (367.719, 949.950)\\
			&			$\beta_{0}^*$ & -8.069 & (-8.177, -7.957)\\
			&			$\beta_{1}^*$ & 0.009 & (0.006, 0.011)\\
			\hline 
			BYM & $\tau^2$ &  0.108 & (0.012, 0.470) \\
			&			$\nu^2$ & 0.023 & (0.003, 0.173)\\
			&			$\beta_{0}^*$ & -7.917 & (-8.167, -7.694)\\
			&			$\beta_{1}^*$ & 0.007 & (0.001, 0.014)\\
			\hline 
			LGCP & $\sigma^2$ &  0.479 & (0.237, 0.914) \\
			&			$\phi$ & 1163.877 & (528.618, 1967.756)\\
			&			$\beta_{0}$ & -19.333 & (-19.738, -19.013)\\
			&			$\beta_{1}$ & 0.008 & (0.001, 0.015)\\
			\hline
		\end{tabular}
	\end{center}
\end{table}

\end{document}